\documentclass[prl,aps,twocolumn,showpacs,nobibnotes,epsf]{revtex4}

\usepackage{graphicx}
\usepackage{dcolumn}
\usepackage{bm}
\usepackage{SIunits}

\begin{document}
\title{Phase diagram as a function of doping level and pressure in Eu$_{1-x}$La$_x$Fe$_2$As$_2$ system}
\author{M. Zhang$^1$, J. J. Ying$^1$, Y. J. Yan$^1$, A. F. Wang$^1$, X. F. Wang$^1$, Z. J. Xiang$^1$,
G. J. Ye$^1$, P. Cheng$^1$, X. G. Luo$^1$, Jiangping Hu$^{2,3}$ and
X. H. Chen$^1$} \altaffiliation{Corresponding author}
\email{chenxh@ustc.edu.cn} \affiliation{1. Hefei National Laboratory
for Physical Science at Microscale and Department of Physics,
University of Science and Technology of China, Hefei, Anhui 230026,
People's Republic of
China\\
2.  Beijing National Laboratory for condensed Matter Physics,
Institute of Physics, Chinese Academy of Sciences, Beijing 100190,
China\\ 3. Department of Phyiscs, Purdue University, West Lafayette,
Indiana 47907, USA}

\begin{abstract}
We establish the phase diagram of Eu$_{1-x}$La$_x$Fe$_2$As$_2$
system as a function of doping level x and the pressure by measuring
the resistivity and magnetic susceptibility. The pressure can
suppress the spin density wave (SDW) and structural transition very
efficiently,  while enhance the antiferromagnetic (AFM) transition
temperature T$_N$ of Eu$^{2+}$. The superconductivity coexists with
SDW order at the low pressure, while always coexists with the
Eu$^{2+}$ AFM order. The results  suggests that Eu$^{2+}$ spin
dynamics is disentangeld with  superconducting (SC)  pairing taken
place in the two-dimensional \emph{Fe-As} plane, but it can strongly
affect superconducting coherence  along c-axis.
\end{abstract}


\vskip 300 pt

\maketitle

Iron-based superconductors have attracted great attentions these
years\cite{Kamihara,chenxh, ren, rotter}. The parent compound
undergoes structural and spin density wave (SDW) transitions. With
chemical doping or high pressure, both structural and SDW transition
can be suppressed and superconductivity emerges. $A$Fe$_2$As$_2$
($A$=Ca, Sr, Ba, Eu) with the $ThCr_2Si_2$-type structure were
widely investigated because it is easy to grow large size of high
quality single crystals\cite{wang}. The maximum $T_C$ for the
hole-doped samples is about 38 K and for the Co doped samples the
maximum $T_C$ is about 26 K\cite{rotter, Sefat}. Recently,
superconductivity up to 49 K was discovered in rare earth doped
CaFe$_2$As$_2$, it is the highest T$_C$ observed in 122
system\cite{Saha, Gao, Lv, Qi}. EuFe$_2$As$_2$ has the same
structure as that of CaFe$_2$As$_2$\cite{ren1}. EuFe$_2$As$_2$ shows
superconductivity around 31 K under the pressure of 3
GPa\cite{Kurita}.

The Eu122 system has an intriguingly outstanding issue regarding the
interplay between the magnetism of large  Eu$^{+2}$ spins and the
superconductivity. The doping on the Eu122 system has been achieved
in two ways:  direct electron doping on Fe-As layer, such as
Co-doped Eu122,  and   doping induced by $Eu$ replacement with rare
earth atoms such as Eu$_{1-x}$La$_x$Fe$_2$As$_2$. In the former
case, a resistivity reentrance due to AFM order of Eu$^{+2}$ spins
was often observed \cite{ren2, Miclea, Ying, He}, an indication of
the effect of Eu$^{2+}$ spins on superconductivity. However, a clear
understanding of such an effect has not been established due to
lacking of systematic investigation.  In the latter case, earlier
results on Eu$_{1-x}$La$_x$Fe$_2$As$_2$ shows only gradual decrease
of the spin density wave transition temperature, T$_{SDW}$, of $Fe$
spins with increasing the La content. No superconducting transition
was observed \cite{wu} and no information regarding of the role of
Eu$^{+2}$ spins has been provided.

High pressure is confirmed to be good method to tune the phase
competition of superconductivity and SDW in iron-pnicides
superconductors. In this letter, we systematically investigate the
superconductivity in La doped EuFe$_2$As$_2$ system under ambient
and high pressure by resistivity and magnetic susceptibility
measurements. We obtain a complete phase diagram. obtained. It is
found that one superconducting transition was observed in the
highest doping level La-doped EuFe$_2$As$_2$ sample, while  the zero
resistivity cannot be reached under ambient pressure. Both
structural and SDW transitions can be completely suppressed and zero
resistivity can be achieved under high pressure. The intriguing
result is that the antiferromagnetic (AFM) transition temperature
T$_N$  of the local moments of the Eu$^{2+}$ ions is always higher
than $T_C$, and is not affected by La doping, while slightly
increases with increasing the pressure. It suggests that although
the superconductivity always coexists with AFM order of Eu$^{2+}$
ions, the two orders are disengaged with each other in La doped
EuFe$_2$As$_2$. Together with earlier results on Co-doped Eu122
\cite{ren2, Miclea, Ying, He}, these results suggest that Eu$^{2+}$
spins only strongly affect SC coherence along c-axis, but not the SC
pairing in \emph{Fe-As} layers.

High quality single crystals with nominal composition
Eu$_{1-x}$La$_x$Fe$_2$As$_2$(x=0, 0.15, 0.3, 0.4, 0.5) were grown by
conventional solid-state reaction using FeAs as self-flux. Clean Eu
bulk, Fe powder and As powder were employed as starting materials.
Starting materials were weighed in the stoichiometric ratio inside
an Ar-filled glove box. The mixture was loaded into an alumina
crucible and then sealed into a quartz tube under vacuum. It was
slowly heated to 680 $\celsius$, held for 12h, and then heated to
1170 $\celsius$ and heated for 10 h, and then the quartz tube was
cooled to 900 $\celsius$ at a rate of 4 K/h. Finally, the quartz
tube was cooled in the furnace after shutting off the power. The
shining plate-like Eu$_{1-x}$La$_x$Fe$_2$As$_2$ crystals were
mechanically cleaved from the flux and obtained for measurements.
The Energy-dispersive X-ray spectroscopy (EDX) indicates that the
actually doping x was 0, 0.08, 0.15, 0.22 and 0.27 for these five
La-doping samples, respectively. It has to be addressed that the
maximum solid solution x cannot be achieved beyond 0.27 although the
nominal doping level is much larger than 0.5. Pressure was generated
in a Teflon cup filled with Daphne Oil 7373 which was inserted into
a Be-Cu pressure cell. The resistivity measurements were performed
using the {\sl Quantum Design} PPMS-9, and magnetic susceptibility
was measured using the Quantum Design SQUID-MPMS.
\begin{figure}[t]
\centering
\includegraphics[width=0.5\textwidth]{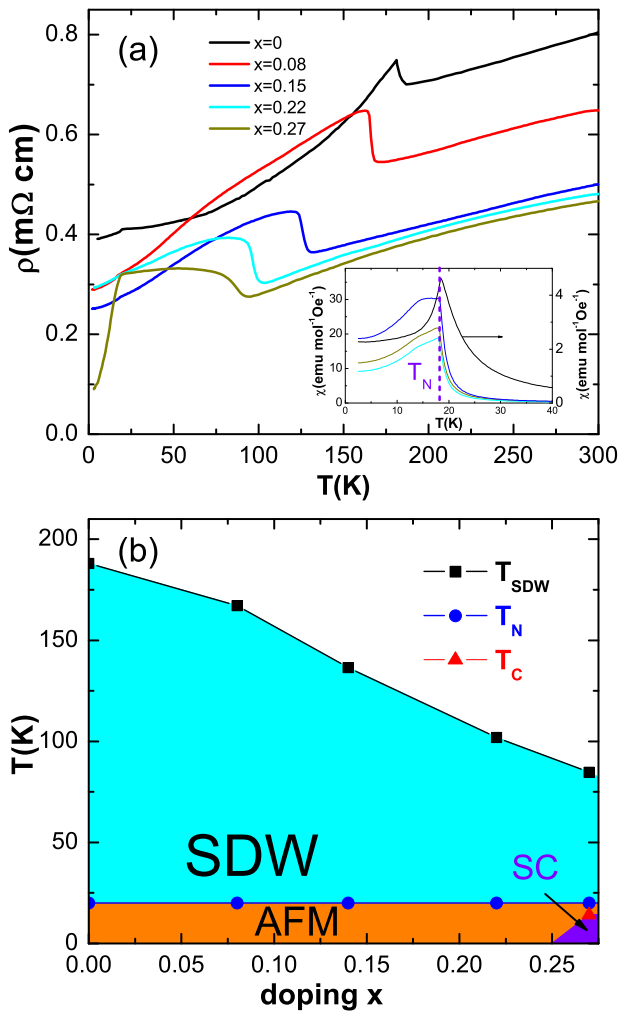}
\caption{(color online). (a): Temperature dependence of resistivity
for Eu$_{1-x}$La$_x$Fe$_2$As$_2$ with different doping x. The inset
shows the temperature dependence of susceptibility below 40 K. (b):
The phase diagram of Eu$_{1-x}$La$_x$Fe$_2$As$_2$ derived from
resistivity and susceptibility. The AFM indicates the
antiferromagnetic state of Eu$^{2+}$.}\label{fig1}
\end{figure}

Figure 1(a) shows the temperature dependence of resistivity for the
samples of Eu$_{1-x}$La$_x$Fe$_2$As$_2$ with different doping
levels. For the parent compound, two anomalies in resistivity are
observed at 188 K and 20 K, respectively. They arise from the
SDW/structural transitions and the AFM order of local moments for
$Eu^{2+}$ ions\cite{wu}. The resistivity anomaly arose from the the
SDW/structural transition is gradually suppressed with increasing
the doping x, while the anomaly around 20 K due to the
antiferromagnetic transition of Eu$^{2+}$ ions nearly does not
change by varying the La doping. The inset shows the temperature
dependence of susceptibility for the samples of
Eu$_{1-x}$La$_x$Fe$_2$As$_2$ with different x.  It is found that the
antiferromagnetic transition temperature (T$_N$) does not change
although the content of Eu$^{2+}$ decreases by substitution of the
La$^{3+}$. Such behavior in susceptibility is consistent with that
observed in the resistivity. A superconducting transition is
observed in the sample of x = 0.27, however, the zero resistivity
cannot reach. Figure 1(b) shows the phase diagram obtained from the
resistivity and susceptibility measurements for the
Eu$_{1-x}$La$_x$Fe$_2$As$_2$ system. The structural/SDW transition
temperature is suppressed to about 85 K, and superconductivity
emerges when La doping is increased to 0.27. However, solid solution
x for the Eu$_{1-x}$La$_x$Fe$_2$As$_2$ cannot be larger than 0.27.
Therefore, we cannot obtain a whole phase diagram under the ambient
pressure in the Eu$_{1-x}$La$_x$Fe$_2$As$_2$ system. The phase
diagram shows the $T_N$ is nearly independent of the La doping.
\begin{figure}[t]
\centering
\includegraphics[width=0.5\textwidth]{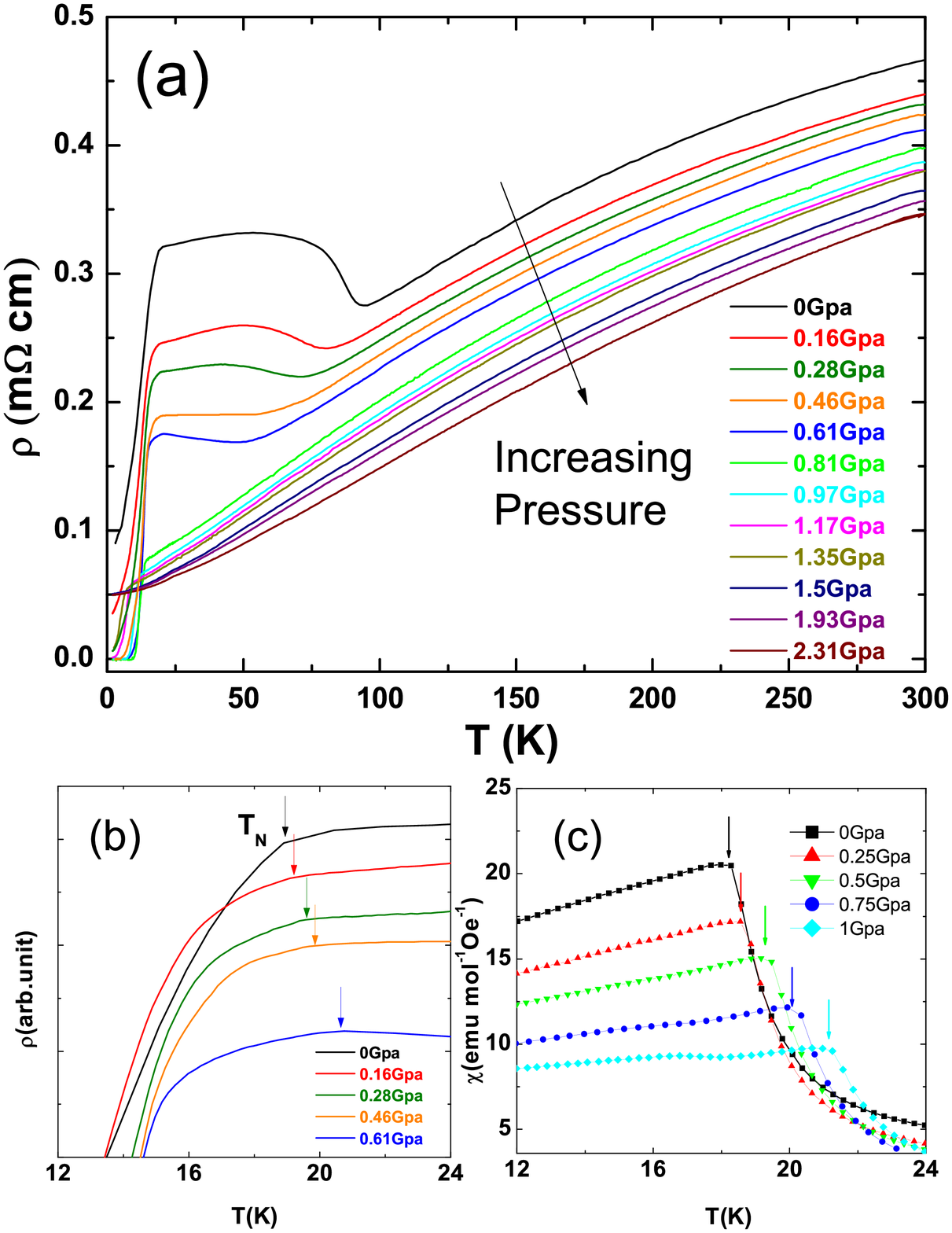}
\caption{(color online). (a): Temperature dependence of resistivity
for Eu$_{0.73}$La$_{0.27}$Fe$_2$As$_2$ under different pressures.
(b): The enlarged area of the temperature dependent resistivity
under various pressures around T$_N$. (c): The enlarged area of the
temperature dependence of susceptibility under various pressures
around T$_N$.} \label{fig2}
\end{figure}

\begin{figure}[t]
\centering
\includegraphics[width=0.5\textwidth]{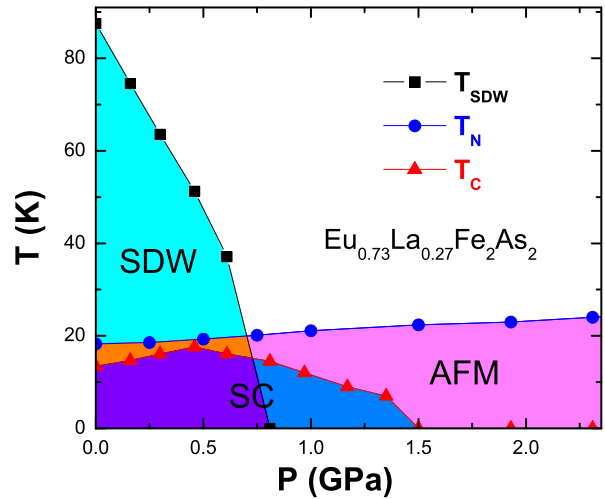}
\caption{(color online). The pressure phase diagram of
Eu$_{0.73}$La$_{0.27}$Fe$_2$As$_2$ derived from resistivity and
susceptibility. The AFM indicates the antiferromagnetic state of
Eu$^{2+}$. } \label{fig3}
\end{figure}

\begin{figure}[t]
\centering
\includegraphics[width=0.5\textwidth]{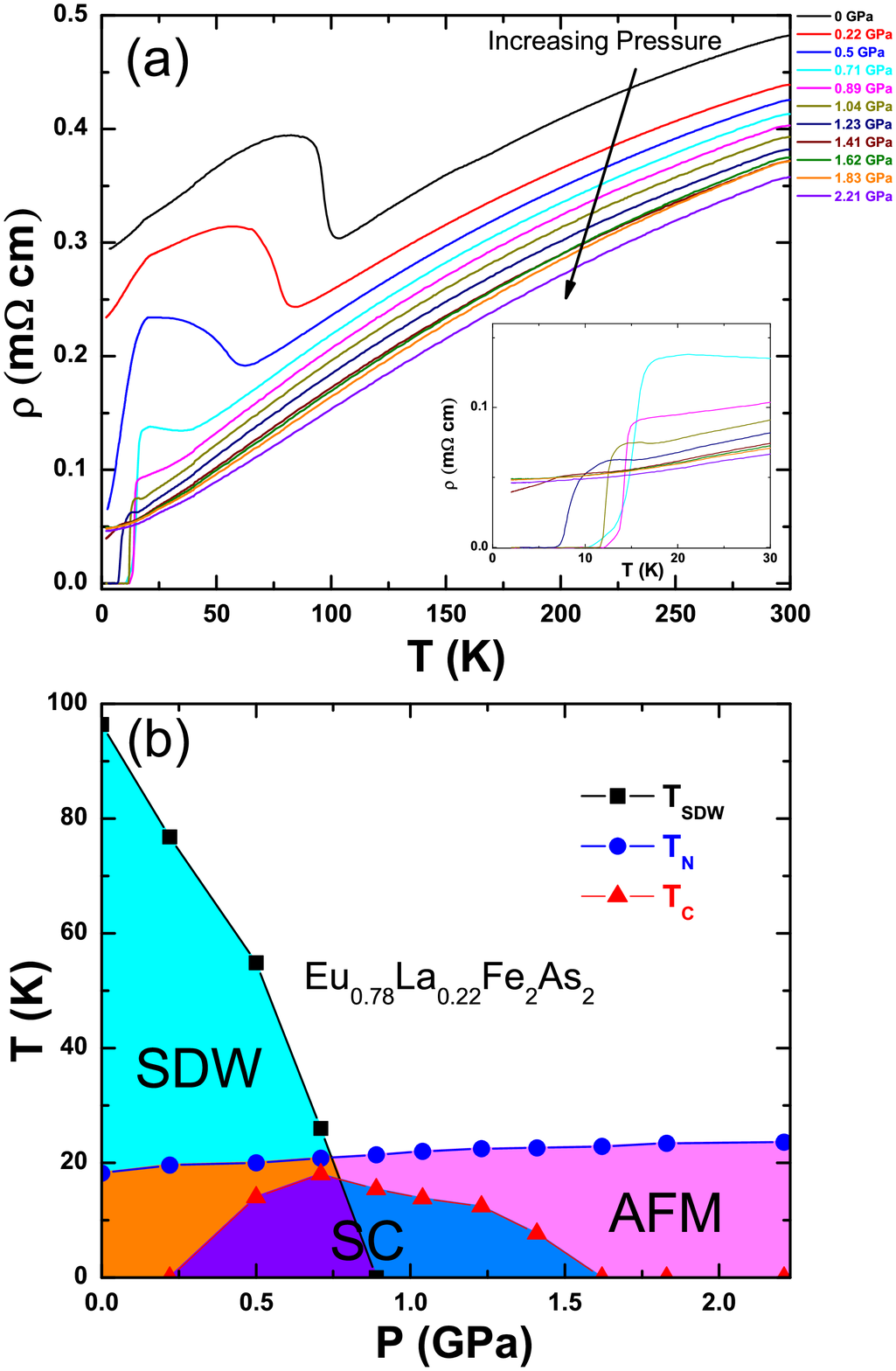}
\caption{(color online). (a): Temperature dependence of resistivity
for Eu$_{0.78}$La$_{0.22}$Fe$_2$As$_2$ under different pressures.
The inset is the enlarged area of resistivity around $T_C$. (b): The
pressure phase diagram of Eu$_{0.78}$La$_{0.22}$Fe$_2$As$_2$ derived
from resistivity and susceptibility. The AFM indicates the
antiferromagnetic state of Eu$^{2+}$. } \label{fig4}
\end{figure}

In order to obtain a complete phase diagram in the
Eu$_{1-x}$La$_x$Fe$_2$As$_2$ system, we performed resistivity and
susceptibility measurement for the sample x = 0.27 with the maximal
La doping under the hydrostatic pressure. It is well known that the
pressure provides another way to suppress SDW order and to induce
the supercodncutivity, therefore, the complete phase diagram is
expected under the pressure in the Eu$_{1-x}$La$_x$Fe$_2$As$_2$
system. Figure 2(a) shows the temperature dependence of resistivity
for the sample with x=0.27. It is found that the pressure can
suppress both structure and SDW transition very efficiently and
greatly improves the superconductivity. Zero resistivity is achieved
under the pressure of 0.46 GPa. With increasing the pressure larger
than 0.81 GPa, the structure and SDW transitions are completely
suppressed. When the applied pressure larger than about 1.5 GPa, the
superconductivity disappears. As shown in Fig.2(b), the temperature
corresponding to the kink due to the AFM order of Eu$^{2+}$ ions in
resistivity increases with increasing the pressure. To confirm the
behavior, we also measured the susceptibility with the magnetic
field applied within the ab-plane of the crystal under pressure.
Figure 2(c) shows the temperature dependence of susceptibility under
different pressure. It clearly shows that the antiferromagnetic
transition slightly increases with increasing the pressure, being
consistent with the results from the resistivity measurement.

Based on the results of resistivity and susceptibility under
pressure as shown in Fig.2, a complete phase diagram is obtained,
and plotted in Fig.3. The phase diagram is very complicated and
similar to that of $Ba_{1-x}K_xFe_2As_2$ and
$BaFe_{2-x}Co_xAs_2$\cite{chen,wang1}. As shown in the phase
diagram, the Eu$_{0.73}$La$_{0.27}$Fe$_2$As$_2$ system shows the
coexistence of superconductivity and SDW under the pressure less
than 0.8 GPa. Besides, there exists an antiferromagnetic order arose
from the local moment of $Eu^{2+}$ ions. Such AFM order always
occurs in the various applied pressures. It should be pointed out
that T$_N$ is always higher than the $T_C$. It indicates that the
three phases of superconductivity, SDW and AFM order coexist under
the pressure less than 0.8 Gpa.  In this sense, the phase diagram is
complicated relative to those of $Ba_{1-x}K_xFe_2As_2$ and
$BaFe_{2-x}Co_xAs_2$. The superconductivity is completely suppressed
under the pressure of 1.5 Gpa much less than that in the parent
compound EuFe$_2$As$_2$. It suggests that the La-doping and pressure
have the same effect on the superconductivity and on SDW. However,
the pressure can enhance the $T_N$.

We also studied the phase diagram of the sample
Eu$_{0.78}$La$_{0.22}$Fe$_2$As$_2$ by measuring resistivity and
susceptibility. Figure 4(a) shows the temperature dependence of
resistivity under the various pressure.
Eu$_{0.78}$La$_{0.22}$Fe$_2$As$_2$ shows structural and SDW
transitions around 96 K and no superconducting transition is
detected under the ambient pressure. When applying a small pressure
of about 0.5 GPa, resistivity shows trace of superconducting
transition, being similar to the case for the sample x=0.27 under
the ambient pressure. T$_C$ gradually increases and zero resistivity
is achieved with further increasing the pressure. When the applied
pressure larger than 0.7 GPa, T$_C$ starts to decrease with
increasing the pressure, and superconductivity is completely
suppressed when applied pressure up to 1.62 GPa. The resistivity
behavior of Eu$_{0.78}$La$_{0.22}$Fe$_2$As$_2$ under small pressure
is very similar to that observed in the sample of
Eu$_{0.73}$La$_{0.27}$Fe$_2$As$_2$ at ambient pressure. It further
indicates that La doping has the same effect on SDW and
superconductivity as the applied pressure. The phase diagram of the
sample Eu$_{0.78}$La$_{0.22}$Fe$_2$As$_2$ as a function of pressure
is plotted in Fig.4(b). The phase diagram is the same as that with
the control parameter of x observed in $Ba_{1-x}K_xFe_2As_2$ and
$BaFe_{2-x}Co_xAs_2$ system except for the existence of AFM order of
$Eu^{2+}$ ions.

Both structural and SDW transition are suppressed with La doping,
and only superconducting transition can be achieved and no
zero-resistance is reached in the Eu$_{1-x}$La$_x$Fe$_2$As$_2$
system. The AFM transition of Eu$^{2+}$ ions is barely affected by
La-doping.  With applying an external pressure, the
superconductivity can be improved drastically on the highly
La-doping samples. The SC dome appears in the T-P phase. However,
the AFM transition of  Eu$^{2+}$ is hardly affected by the pressure.
These results suggest that the superconductivity is disentangled
with the AFM order in the Eu$_{1-x}$La$_x$Fe$_2$As$_2$ system .

It is intriguing to compare our results with earlier results on other Eu122 compounds\cite{He, Park,
Alireza, Hu}. In the parent compound
 EuFe$_2$As$_2$, T$_C$ is about 31K under pressure.  A resistivity reentrance
due to the antiferromagnetic ordering of Eu$^{2+}$ spins is widely
observed in the other Eu122 systems below T$_C$\cite{He, Park,
Alireza}.  In Sr$_{1-y}$Eu$_y$(Fe$_{0.88}$Co$_{0.12}$)$_2$As$_2$ single
crystals\cite{Hu}, Hu et al. claimed that the bulk SC disappears
suddenly when T$_N$ $>$ T$_C$. These results definitely indicates the strong competition between the AFM
and SC.  Thus, we reach a dilemma  regarding the relation between the AFM and SC in such an comparison.

This dilemma can be easily resolved if we assume the major effect of  Eu$^{2+}$ spins on SC
is along c-axis. In a quasi-two dimensional system,  a global SC state is only formed when the
SC coherence is reached along c-axis. In  La-doped Eu122,  the  doped electrons to Fe-As layer
stems from  La.  The SC coherence along c-axis can be achieved by virtual hopping through La atoms,
which minimizes the effect of  the dynamics of Eu$^{2+}$ spins.  However, in the parent compounds or
Co-doped Eu122,  the similar processes go through Eu and Eu$^{2+}$ spins can strongly affects the SC
coherence along c-axis.  If the pairing in Fe-As is strong, the resistivity reentrance  can take
place due to the block of c-axis coherence by  the critical fluctuations of Eu$^{2+}$ spins near the
AFM transition, which explains the observed phenomena in the parent compounds of Eu122.
If the pairing in Fe-As is weak, the phase coherence along c-axis may be never developed under
the influence of Eu$^{2+}$ spin, which accounts for what observed in the Co-doped Eu122.

In conclusion, we established the complete phase diagram by
measuring the resistivity and susceptibility under various pressure
for the crystals of Eu$_{1-x}$La$_x$Fe$_2$As$_2$ with different x.
Only trace of superconductivity was observed in the highly La-doped
samples under ambient pressure. High pressure efficiently suppresses
the SDW and structure transition, and improves the
superconductivity, while leads to an increase in T$_N$. The
superconductivity coexists with SDW order at the low pressure, while
always coexists with the antiferromagnetic order of $Eu^{2+}$ spins.

{\bf ACKNOWLEDGEMENT} This work is supported by the National Natural
Science Foundation of China (Grant No. 51021091), National Basic
Research Program of China (973 Program, Grant No. 2011CB00101 and
No. 2012CB922002) and Chinese Academy of Sciences.


\begin{references}
\bibitem{Kamihara}
Y. Kamihara $et$ $al$., \emph{J. Am. Chem. Soc.} {\bf 130},
3296(2008).
\bibitem{chenxh}
X. H. Chen $et$ $al$., Nature {\bf 453}, 761(2008).
\bibitem{ren}
Z. A. Ren $et$ $al$., Europhys. Lett. {\bf 83}, 17002(2008).
\bibitem{rotter}
M. Rotter $et$ $al$., Phys. Rev. Lett. {\bf 101}, 107006(2008).
\bibitem{wang}
X. F. Wang $et$ $al$.,  Phys. Rev. Lett. {\bf 102}, 117005(2009).
\bibitem{Sefat}
Athena S. Sefat $et$ $al$., Phys. Rev. Lett. {\bf 101}, 117004
(2008).
\bibitem{Saha}
S. R. Saha $et$ $al$., arXiv:1105.4798
\bibitem{Gao}
Zhaoshun Gao $et$ $al$., EPL, {\bf 95}, 67002 (2011).
\bibitem{Lv}
B. Lv $et$ $al$., Proc. Nat. Acad. Sci. {\bf 108}, 15705 (2011).
\bibitem{Qi}
Yanpeng Qi $et$ $al$., arXiv:1106.4208
\bibitem{ren1}
Zhi Ren $et$ $al$., Phys. Rev. B {\bf 78}, 052501 (2008).
\bibitem{Kurita}
Nobuyuki Kurita $et$ $al$., Phys. Rev. B {\bf 83}, 214513 (2011).
\bibitem{ren2}
Zhi Ren $et$ $al$., Phys. Rev. Lett. {\bf 102}, 137002 (2009).
\bibitem{Miclea}
 C. F. Miclea $et$ $al$., Phys. Rev. B {\bf 79}, 212509 (2009).
\bibitem{Ying}
J. J. Ying $et$ $al$., Phys. Rev. B {\bf 81}, 052503 (2010).
\bibitem{He}
Y. He $et$ $al$., J. Phys.: Condens. Matter {\bf 22} 235701(2010).
\bibitem{wu}
T. Wu $et$ $al$., J. Mag. Mag. Mat. {\bf 321}, 3870-3874 (2009).
\bibitem{chen}
H. Chen $et$ $al$., EPL, {\bf 85}, 17006 (2009).
\bibitem{wang1}
X. F. Wang $et$ $al$., New J. Phys. {\bf 11}, 045003 (2009).
\bibitem{Park}
Tuson Park $et$ $al$., J. Phys.: Condens. Matter {\bf 20}, 322204
(2008).
\bibitem{Alireza}
Patricia L Alireza $et$ $al$., J. Phys.: Condens. Matter {\bf 21},
012208 (2009).
\bibitem{Hu}
Rongwei Hu $et$ $al$., Phys. Rev. B {\bf 83}, 094520 (2011).


\newpage


\noindent

\end{references}
\end{document}